\documentclass[12pt]{article}
\usepackage[utf8]{inputenc}
\usepackage{empheq}
\usepackage{bm}
\usepackage[normalem]{ulem}
\usepackage{latexsym}
\usepackage{color}
\usepackage{amsmath,amsfonts,amssymb}
\usepackage{graphicx,epsfig}
\usepackage{psfrag}
\usepackage{amsthm}
\usepackage{cite,url,comment}
\usepackage{makecell}
\usepackage{dsfont}

\pdfoutput=1

\usepackage{mathrsfs}
\usepackage{fullpage}
\usepackage{setspace}
\usepackage{bm}
\usepackage{bbm}

\usepackage[normalem]{ulem}
\usepackage{enumerate}
\usepackage{yfonts}

\usepackage{psfrag}

\usepackage{cancel}
\usepackage{slashed}

\usepackage[font=small,labelfont=bf]{caption}
\usepackage{subcaption}

\usepackage[colorlinks=true]{hyperref} 

\newcommand{\be}{\begin{equation}}
\newcommand{\bea}{\begin{eqnarray}}
\newcommand{\ee}{\end{equation}}
\newcommand{\eea}{\end{eqnarray}}
\numberwithin{equation}{section}

\hypersetup{colorlinks=false, linkcolor=blue, citecolor=red}
	\def\be{\begin{equation}}
	\def\ee{\end{equation}}

	\newcommand{\vs}[1]{\vspace{#1 mm}}
	
\begin{document}
	\begin{flushright}
		%
	\end{flushright}
	\begin{center}
		{\Large{\bf Complexity of warped conformal field theory}}\\

\vs{10}

{\large
Arpan Bhattacharyya${}^{a,\,}$\footnote{\url{abhattacharyya@iitgn.ac.in}}, Gaurav Katoch${}^{b,\,}$\footnote{\url{gauravitation@gmail.com}}, Shubho R. Roy${}^{b,\,}$\footnote{\url{sroy@phy.iith.ac.in}}}

\vskip 0.3in

{\it ${}^{a}$ Indian Institute of Technology, Gandhinagar, Gujarat-382355, India}\vskip .5mm

{\it ${}^{b}$ Department of Physics,
Indian Institute of Technology Hyderabad,\\
Kandi, Sangareddy, Telengana 502285, India}\vskip .5mm

\vskip.5mm

\end{center}

\vskip 0.35in

\begin{abstract}
Warped conformal field theories in two dimensions are exotic nonlocal, Lorentz violating field theories characterized by Virasoro-Kac-Moody symmetries and have attracted a lot of attention as candidate boundary duals to warped AdS$_3$ spacetimes, thereby expanding the scope of holography beyond asymptotically AdS spacetimes. Here we investigate WCFT$_2$\,s using \emph{circuit complexity} as a tool. First we compute the holographic volume complexity (CV) which displays a linear UV divergence structure, more akin to that of a local CFT$_2$ and has a very complicated dependence on the Virasoro central charge $c$ and the $U(1)$ Kac-Moody level parameter $k$. Next we consider circuit complexity based on Virasoro-Kac-Moody symmetry gates where the complexity functional is the geometric (group) action on coadjoint orbits of the Virasoro-Kac-Moody group. We consider a special solution to extremization equations for which complexity scales linearly with ``time''. In the semiclassical limit (large $c,k$, while $c/k$ remains finite and small) both the holographic volume complexity and circuit complexity scales with $k$.
	\end{abstract}

	\newpage

\tableofcontents

\section{Introduction}

	Holography \cite{Maldacena:1997re, Gubser:1998bc, Witten:1998qj, Aharony:1998ub} has not only provided us with tools which have revolutionized our understanding of phenomena in strongly coupled field theories, it has even led to the discovery of novel exotic phases of strongly coupled field theories and led to the identification of new conformal field theories. One such example are the Warped conformal field theories (WCFT) \cite{Hofman:2011zj,Detournay:2012pc}, which are the proposed holographic duals of warped AdS$_3$ spacetimes \cite{Anninos:2008fx}. WCFTs can be defined as the two dimensional field theories with $SL(2,\mathbb{R})_{R}\times U(1)_{L}$ Kac-Moody symmetry, which is the local extension of the algebra of two global translation and one global chiral scale symmetries. This is in contrast to the much older result \cite{Polchinski:1987dy} where an unitary two dimensional QFT with global Poincar\'e and scale invariance, 
	\begin{align*}
x^{-} & \rightarrow x^{-}+a,\qquad x^{+}\rightarrow x^{+}+b,\\
x^{-} & \rightarrow\lambda_{-}\,x^{-},\qquad x^{+}\rightarrow\lambda_{+}x^{+}.
\end{align*}
	ends up having an enhanced to a direct product of two copies of the Virasoro algebra, corresponding to two dimensional conformal symmetry,
	\begin{align*} 
x^{-} & \rightarrow f(x^{-}),\qquad x^{+}\rightarrow g(x^{+}).
\end{align*}
	if the dilatation operator has discrete non-negative spectrum. Here $x^\pm$ are the two dimensional lightcone coordinates. In \cite{Hofman:2011zj}, however the field theory was assumed to possess only one-sided (chiral) global scale invariance	\begin{align*}
x^{-} & \rightarrow x^{-}+a,\qquad x^{+}\rightarrow x^{+}+b,\\
x^{-} & \rightarrow\lambda_{-}\,x^{-}.
\end{align*}
There is a novel symmetry enhancement when one adds to the mix a chiral boost symmetry $x^+\rightarrow x^+ + \omega\, x^-$. In such a case the symmetry algebra gets enhanced to an infinite dimensional symmetry algebra, namely that of a semidirect product of a Virasoro algebra and a $U(1)$ current algebra (Virasoro-Kac-Moody algebra), corresponding to the so called \emph{warped conformal symmetry},
	\begin{align}
x^{-} & \rightarrow f(x^{-}),\qquad x^{+}\rightarrow x^{+}+g(x^{-}) \label{coordtrans}.
\end{align}
\par
Field theories possessing such a warped conformal symmetry are the WCFTs. Since then warped conformal symmetry and WCFTs have been explored intensely using various field theory and holographic tools - it is worth nothing a few prominent works here. See reference \cite{Detournay:2012pc}, for a discussion about representations of the warped conformal symmetry and an analogue of the Cardy formula.  Correlation functions have been worked out in reference  \cite{Song:2017czq}. Two and three point functions get completely determined by the global warped conformal symmetry, while the four-point functions are determined up to an arbitrary function of the cross ratio. 
Several concrete examples of WCFTs have now been worked out, see \cite{Compere:2013aya, Jensen:2017tnb} for bosonic WCFTs, \cite{Hofman:2014loa, Castro:2015uaa, Chaturvedi:2018uov, Davison:2016ngz} for fermionic WCFT models and \cite{Chen:2020juc} for supersymmetric WCFTs. For other interesting works in WCFTs refer to \cite{Jensen:2017tnb, Apolo:2018qpq,Apolo:2018eky,Song:2019txa}. In this work we are particularly interested in holographic WCFTs, which are dual field theory candidates to gravitational theories on warped AdS$_3$ (WAdS$_3$) spacetimes. WAdS$_3$ are non-Einstein spacetimes which can be realized in topologically massive gravity \cite{Vuorio:1985ta, Percacci:1986ja, Ortiz:1990nn, Nutku:1993eb, Gurses:1994bjn} or in string theory \cite{Anninos:2008qb, Compere:2008cw, Detournay:2005fz, Israel:2004vv, Israel:2003ry}. The asymptotic symmetry group of these spacetimes is the semidirect product of a Virasoro algebra and a $U(1)$ affine Kac-Moody algebra \cite{Compere:2007in,Compere:2008cv, Blagojevic:2008bn, Compere:2009zj, Henneaux:2011hv}. These spacetimes are not asymptotically locally AdS, and hence they expand the scope of holography beyond asympotically AdS. In particular we are interested in spacelike warped AdS$_3$ spacetimes, which are obtained when a spatial line or circle is fibered over AdS$_2$. Timelike and null warped AdS$_3$, where the $U(1)$ fiber is timelike and null respectively, are known to contain closed timelike curves (CTC) and hence are not expected to have sensible, well-behaved boundary duals. Spacelike warped AdS$_3$ spacetimes also admit black hole solutions \cite{Anninos:2008fx}.
\par 
Ideas from quantum information have brought new insights into various physics branches and had far-reaching consequences. It has given a new perspective in interpreting several geometric objects in the context of holography. A most studied information-theoretic tool is entanglement entropy. Typically, the entanglement entropy is computed using the von-Neumann entropy after partitioning the systems into two subsystems and tracing them out. This has been extensively explored in the context of AdS/CFT \cite{Rangamani:2016dms} and the Warped holography \cite{Castro:2015csg,Song:2016gtd,Chen:2019xpb,Anninos:2013nja,Apolo:2018oqv,Detournay:2020vrd}. Another information-theoretic quantity, primarily motivated by recent developments in black hole physics \cite{susskind2,Susskind:2014rva}, has come into the limelight is {\sl circuit complexity}\cite{2006Sci...311.1133N,2005quant.ph..2070N}. In the context of holography, certain geometrical objects, e.g. maximal volume of a particular codimension-one bulk slice (complexity = volume) \cite{Stanford:2014jda}, gravitational action defined on Wheeler-DeWitt patch of a bulk Cauchy surface anchored at a specific time (complexity = action) \cite{Brown:2015bva}, are conjectured to be the gravity dual to the circuit complexity of the dual field theory state. Circuit Complexity, an idea from the theory of quantum computation, basically quantifies the minimal number of operations or gates required to build a circuit that will take one from a given reference state($|\psi_R\rangle$) to the desired target state ($|\psi_T\rangle$). In recent times, circuit complexity has been explored in the context of quantum field theory \cite{Jefferson,Chapman:2017rqy, Hackl:2018ptj,Khan:2018rzm,Bhattacharyya:2018bbv,Bhattacharyya:2019kvj,me1,Magan:2018nmu,Caputa:2018kdj,Erdmenger:2020sup,Flory:2020eot,Flory:2020dja,Erdmenger:2021wzc,Chagnet:2021uvi,Koch:2021tvp} \footnote{This list is by no means exhaustive. Interested readers are referred to these reviews \cite{Chapman:2021jbh,Bhattacharyya:2021cwf}, and citations are therein for more details.}. In this paper, we will explore complexity both from the field theory and gravity side in the context of warped holography, complementing the studies of entanglement entropy in this context.

\par The plan of the paper is as follows. In section \ref{CVV} we resort to holographic methods, in particular the complexity-volume (CV) prescription to compute the complexity of the warped conformal field theories dual to timelike and spacelike warped AdS$_3$ spacetimes realized in topologically massive gravity theory. We find that for the timelike case, the dependence on the UV cutoff is rather complicated - an indication of the fact the warped CFT is a nonlocal theory, but the holographic complexity is not defined for a arbitrary values of the UV cutoff. The complexity is only well defined when the cutoff is kept under a critical value determined by the warping parameter. Such a phenomenon has already been observed in the case of complexity of field theories dual to null warped AdS$_3$ spacetimes realized in the context of $T\overline{T}, J\overline{T}, \overline{J}T$ deformed CFT$_2$'s in a different work \cite{Katoch:2022hdf}. Such computations lend credence to the claim that the warped CFT$_2$'s  which are dual to null and timelike warped AdS$_3$ spacetimes do not have an unitary UV completion. Then we work out the holographic complexity of WCFTs dual to spacelike WAdS$_3$ spacetimes.  These are free from pathologies (i.e. are unitary and UV complete) and the underlying symmetry structure is that of a semidirect product of a Virasoro and a $U(1)$ Kac-Moody algebra. The holographic complexity in this case scales extensively with system in units of the UV cutoff, a trait which is perhaps more expected from a local CFT$_2$, despite the fact that warped CFTs are highly nonlocal theories. There is a nontrivial dependence on the symmetry parameters $c,k$. In particular the complexity \emph{does not} scale linearly with the Virasoro central charge $c$ as it did in the case of local CFT$_2$, but instead with the $U(1)$ Kac-Moody level number, $k$. Although there is no restriction on the cutoff in terms of the warping parameter, it can be arbitrary, the complexity is still defined in a restricted domain of the parameter space of the symmetry algebra, namely $c/k\leq25/8$. Next in section \ref{KacMoody}, we adopt the method of \cite{Caputa:2018kdj,Flory:2020eot, Erdmenger:2020sup} to directly evaluate the circuit complexity of WCFT employing the Kac-Moody symmetry gates to construct a complexity functional. Although finding the most general solution to the extremization conditions of this complexity functional appears highly intractable, we are able to obtain an exact solution by simple inspection.  For this special solution, we find that that the complexity has an overall linear scaling in the warping parameter $k$, and has a subleading contribution of order $c/k$. This dependence on $k$, $\frac{c}{k},$ is in good agreement with what we obtain from the gravity side. We conclude the section after an elaborate comparison with the results coming from holography. Finally in section \ref{discussion}, we discuss our results and provide an outlook for further future investigations. Some of mathematical details are provided in appendices \ref{App.A}, \ref{App.B} and \ref{App.C}.

We note that there have been other, complementary studies of complexity of warped conformal field theories \cite{Ghodrati:2019bzz} as well holographic complexity of warped AdS$_3$ black holes \cite{Ghodrati:2017roz,Auzzi:2018pbc,Auzzi:2019fnp}.

\section{Holographic Complexity of warped CFTs}\label{CVV}
	
In this section our goal is to study the complexity of two dimensional WCFTs using holography i.e. using the dual warped-$AdS_3$ solutions. To be precise we use the holographic volume complexity prescription \cite{Susskind:2014rva,susskind2,Stanford:2014jda}. Although timelike and null warped AdS$_3$ spacetimes are not supposed to be dual to any UV complete boundary field theory we work out the holographic complexity of timelike WAdS$_3$ for the sake of completeness. The holographic volume complexity expression must exhibit a characteristic signature for the sickness of the boundary dual field theory. The holographic of null warped AdS$_3$ has already been considered elsewhere \cite{Katoch:2022hdf} where it is obtained as the holographic dual to a little string theory (LST) obtained as a single trace $T\bar{T}, J\bar{T},\bar{J}T$ deformation of a CFT$_2$, for a very special case of the deformation parameters ($\lambda=\epsilon_+=0$). There it has been observed that both the holographic volume and action complexity expressions become either complex or ill-defined if the UV cut off is arbitrarily large. Sensible (real positive) complexity expressions are only obtained when the UV cutoff is restricted by the warping parameter. Such a behavior of the complexity clearly signals the UV incompleteness of the putative WCFT (in this case an LST) dual to the null Warped AdS$_3$.  To avoid redundancy, we skip the null warped case as it has appeared in a separate work \cite{Katoch:2022hdf} and instead we begin our holographic computations with the case of the timelike warped $AdS_3$. We work specifically with the metric in a Poincar\'e  patch of the timelike warped AdS$_3$, which can be obtained by taking the zero temperature limit of the warped black string metric equation (4.10) of \cite{Song:2016gtd} as reviewed in the appendix \ref{App.A}. \\
	The metric in Poincar\`e  patch is (\ref{TWAdS}) in appendix \ref{App.A} and reads
	\begin{align}
		ds^2=\frac{-\text{$dt$}^2+\text{$dx$}^2+\text{$dz$}^2}{z^2}	-\lambda^2\frac{ (\text{$dt$}+\text{$dx$})^2}{4 z^4}\nonumber~.
	\end{align}	
As usual $z=0$ is the (conformal) boundary, and the (warped) AdS radius is set to unity. Here $\lambda$ is a dimensionless parameter representing (timelike) warping. Note that for $z<\lambda/2$, the transverse $x$-direction turns timelike so the conformal boundary is 0+2 dimnsional (two times). From the bulk sting background perspective, the transverse direction $x$ is a compact (closed), and there appears closed timlike curves once one crosses into the deep UV (near boundary) region $2z<\lambda$. Then to obtain a causal macrosopic semiclassical bulk one is forced to excise the spacetime time region $2z<\lambda$. This phenonmenon points out to the UV incompleteness of the warped CFTs dual to timelike warped AdS$_3$ akin to the case of the null warped case. To recap, it can be shown that the dual LST to bulk null warped AdS$_3$ has complex energy eigenvalues for energy scales large $\epsilon$, thereby rendering the dual theory nonunitary \cite{Chakraborty:2020xyz}.  One is forced to truncate the theory beyond a certain cutoff UV energy scales (Hagedorn) such that the spectrum of the truncated theory is real. Although the explicit dual WCFT to the timelike warped AdS$_3$ gravity is as yet unknown one anticipates the dual WCFT to share a similar pathology - the far UV spectrum must be truncated to keep the dynamics unitary. 

In order to compute the volume complexity, we need to first work out the maximal volume spatial slice $\Sigma$ - a spacelike hypersurface which has the maximum volume among all spacelike hypersurfaces anchored at a given boundary time, $T_0$. The volume complexity of the dual boundary theory at the time $T_0$ is then proportional to the volume $\mathcal{V}_\Sigma$ of the maximal volume slice $\Sigma$,
\begin{equation}
\mathcal{C}(T_0 ) = \frac{\mathcal{V}_\Sigma (T_0)}{G_N\, l}\,.
\end{equation}
Here $l$ is some characteristic length scale of the geometry (which is a bit arbitrary to some extent). In the present case we will take it to be the (W)AdS$_3$ radius (which we have set to unity $l=1$).

Let us parameterize a generic spatial surface (say $\gamma$) by $t=t(z)$,$ \forall x$. Then the induced metric on this spacelike hypersurface is
	\begin{align}
		ds^2_{\gamma}&=\left(\frac{1}{z^2}-\frac{\lambda ^2 t'(z)^2}{4 z^4}-\frac{t'(z)^2}{z^2}\right)dz^2-\frac{\lambda ^2 t'(z)}{2 z^4}dx dz+dx^2 \left(\frac{1}{z^2}-\frac{\lambda ^2}{4 z^4}\right) \nonumber~.
	\end{align}
The 
volume of the spacelike hypersurface $\gamma$ is then,
	\begin{align}
		\mathcal{V}_{\gamma}=\int dx\int_{0}^{\infty} dz\,
		\frac{1}{z^2}\sqrt{1-t'(z)^2-\frac{\lambda ^2}{4 z^2}},
	\end{align}
Extremizing this volume functional  leads us to the following Euler-Lagrange equation
	\begin{align}
	z \left(4 z^2-\lambda ^2\right) t''(z)+\left(\lambda ^2-8 z^2\right) t'(z)+8 z^2\, t'(z)^3=0\nonumber~.
	\end{align}
	To solve this differential equation we assume the following ansatz for the spacelike slice anchored at the boundary as
	\begin{align}
		t(z)\text{:=}T_0+T_1 z+T_2 z^2+T_3 z^3+T_4 z^4+T_5 z^5+T_6 z^6+....\nonumber~,
	\end{align}	
	when solved order by order in $z$. Since this is a second order equation, we need a second boundary condition, which is the constraint that asymptotically this is a spacelike surface ($\frac{dt}{dz}|_{z=0}=0$). The solution to the Euler Lagrange equation is remarkably simple, it is the constant time slice $t(z)=T_0$. However plugging in the maximal volume slice $t=T_0$, in the expression for the volume naively gives divergent result since the space is noncompact. So we need to introduce a volume regulator in the form of a radial cutoff, $z=\epsilon$ instead of integrating all the way to the boundary $z=0$. After regulating the volume, one obtains a finite (regulator dependent)\footnote{Note that just like entanglement entropy, complexity is also expected to be a manifestly (UV) cutoff dependent quantity for a continuum quantum field theory. In addition, generically for a quantum theory where states are described by a continuum of state vectors, the so called \emph{circuit complexity} is intrinsically unbounded and can only be defined provided one introduces a tolerance parameter which is a sort of minimal volume cell in the Hilbert space of states. There have been attempts to define quantum complexity which is finite as well as tolerance free but there is no unanimity in such approaches.} expression for the volume complexity of warped CFT dual to a  timelike warped AdS$_3$ spacetime to be
	\begin{align}
		\mathcal{C}&=\frac{1}{G_N }\int dx\int_{\epsilon}^{\infty} dz\,\frac{1}{z^2}\sqrt{1-\frac{\lambda ^2}{4 z^2}}\nonumber\\
		&=\frac{L_x}{ G_N}\left(\frac{1}{\lambda}\sin^{-1}\left(\frac{\lambda}{2\epsilon}\right)+\frac{1}{2\epsilon}\sqrt{1-\frac{\lambda ^2}{4 \epsilon^2}}~\right)\,.	
	\end{align}
	\par
	There are several features to note in this expression for complexity. First and foremost, unlike that of a local CFT$_2$, the complexity of a warped CFT does not diverge linearly with the cutoff $\epsilon$. This is consistent with the fact the warped CFTs are highly nonlocal, boost non-invariant field theories. The second key feature is that for a fixed cutoff, the complexity is a nonanalytic function of the warping parameter $\lambda$ - the complexity does not make sense for all real values of the warping parameter $\lambda$. The cutoff cannot be made arbitrarily small, there is a restriction imposed on it by the warping parameter $\lambda$.  In order for the above complexity expression to make sense, we must always restrict the cutoff to $2\epsilon\geq\lambda$ as we have pointed out earlier that as the radial coordinate, $2z<\lambda$, the bulk $x$-direction turns timlike. Consequently the constant $t$ surface is not a spacelike surface and its volume does not represent complexity. In fact since for $2z<\lambda$, the spacetime turns 1+2-dimensions, no codimension one hypersurface is spacelike and the volume complexity prescription does not apply anymore. This pathology is similar in nature to as the one encounters in the study of null WAdS \cite{Katoch:2022hdf}. Such pathological features render the warped CFTs dual to timelike or null warped AdS$_3$ spacetimes unsuitable for further investigations and in the remainder on will concern ourselves with the warped CFTs which are dual to exclusively spacelike warped AdS$_3.$

	\subsection{Holographic volume complexity of spacelike WAdS$_3$}\label{spacelike}
	Here we consider the physically interesting case of warped CFTs dual to \emph{spacelike} warped AdS$_3$ spacetime. Spacelike warped AdS has isometry group $SL(2, \mathbb{R})\times U(1)$. The metric of spacelike WAdS$_3$ solution is given by \cite{Anninos:2008fx}
	\begin{align}
		ds^2=\frac{l^2}{\nu^2+3}\left[-\cosh ^2\rho ~dt ^2+d\rho ^2+\frac{4 \nu^2 }{\nu^2+3}(dt  ~\sinh \rho +du)^2\right]~.\label{swads3}
	\end{align}
	When $\nu^2>1$ one obtains a spacelike stretched AdS$_3$, while a spacelike squashed AdS$_3$ is obtained when $\nu^2<1$. Evidently $\nu^2=1$ case represents undeformed pure AdS$_3$ spacetime. For computational convenience, we make the diffeomorphism $\tan \theta=\sinh \rho$, with $ 0\leq \theta \leq \pi/2$
	to bring the spacelike warped AdS$_3$ metric to the following form
	\begin{align}
		ds^2=\frac{l^2}{(\nu^2+3) \cos ^2\theta }\left(d\theta ^2-\text{d$t $}^2+\frac{4 \nu^2 }{\nu^2+3}(dt~  \sin \theta +du ~\cos \theta )^2\right)~.\label{conformal}
	\end{align} 
	\par
	As was done previously, the next step towards computing the holographic volume complexity is to locate the maximal volume slice. To this end, let us parameterize a generic spacelike hypersurface by the condition $t=t(\theta)$ $\forall u$. Then the pullback metric on this spacelike hypersurface $t=t(\theta)$ is given by
	\begin{align}
		ds^2=\frac{l^2}{(\nu^2+3) \cos ^2\theta }\left[ -t '(\theta )^2~d\theta ^2+d\theta ^2+\frac{4 \nu^2 }{\nu^2+3}\left( \sin \theta ~ t '(\theta )~d\theta +du ~\cos \theta \right)^2\right]~,
	\end{align}
	with the volume 
	\begin{align}
	\mathcal{V}	=\frac{2\, l^2\,\nu}{(\nu^2+3)^{3/2}}\int du\int d\theta\, \frac{1}{\cos \theta } \sqrt{1-t'^2(
			\theta)}~.\label{volume functional for spacelike WAdS3}
	\end{align}
	Extremizing the volume functional leads to the following Euler-Lagrange equation
	\begin{align}
		-t''(\theta )+\tan \theta ~ t'(\theta )^3-\tan \theta ~ t'(\theta )=0~.\label{EL spacelike warped}
	\end{align}
Apart from the obvious root $t(\theta)= T_0$, this second order nonlinear differential equation has following two nontrivial roots
\begin{align}
t(\theta )=c_2\pm\tan ^{-1}\left(\frac{\sqrt{2} \sin \theta }{\sqrt{2  c_1+\cos 2 \theta +1}}\right)~,\label{null hypersurfaces}
\end{align}
where $ c_2=T_0-\tan ^{-1}\left(\frac{1}{\sqrt{c_1}}\right)$ and $c_1>0$. So these are a pair of one-parameter family (continuum infinity) of codimension one hypersurfaces parameterized by $c_1$. Depending on the value of $c_1$, these could be spacelike ($c_1>1+\frac{\sqrt{\nu ^2+3}}{2 \nu}$), timelike ($c_1<1+\frac{\sqrt{\nu ^2+3}}{2 \nu}$) or null ($c_1=1+\frac{\sqrt{\nu ^2+3}}{2 \nu}$).
A simple inspection of the volume element \eqref{volume functional for spacelike WAdS3} makes it obvious that $t'(\theta)=0$ or $t(\theta)=T_0$ is global maximum among all spacelike slices (refer to Appendix \ref{App.B} for an explicit check). Selecting this constant $t$ spatial slice and evaluating the volume functional, we obtain the holographic complexity of spacelike warped AdS$_3$ is,
	\begin{align}
		\mathcal{C}&= \frac{2 L_xl }{G_N}\frac{\nu}{(\nu^2+3)^{3/2}}\int d\theta\sec\theta~,\nonumber\\
		&=  \frac{2 L_x l}{G_N}\frac{\nu}{(\nu^2+3)^{3/2}} \int_{0}^{1/\epsilon}d\rho~,\nonumber\\
		& = \frac{2 l}{G_N}\frac{\nu}{(\nu^2+3)^{3/2}} \frac{L_x}{\epsilon}~. \label{CV}
	\end{align}
Here we have introduced a radial cutoff $\epsilon$ (boundary UV cutoff) and an IR cutoff, $L_x$, in the transverse boundary direction, $\int du = L_x$ to regulate the complexity expression.

To translate this result in the language of field theory we use the WAdS$_3$/WCFT$_2$ holographic dictionary \cite{Anninos:2008fx,Compere:2009zj,Henneaux:2011hv,Aggarwal:2021zdu}. WAdS$_3$ is realized in topologically massive gravity (TMG) as a classical solution which is asymptotically AdS$_3$ with radius $l$ for every value of the gravitational Chern-Simons (CS) coupling $\mu(>0)$. The CS coupling, $\mu$ is related to the parameter $\nu$ appearing in the gravity solution, $\nu = \frac{\mu l}{3}$. The phase space corresponding to the metric has asymptotic symmetry algebra is a semidirect product of the Virasoro and Kac-Moody algebra with central charge $c$ and the Kac-Moody level number $k$ respectively:
	\begin{align}
     [L_n,L_m]&=(n-m)L_{n+m}+\frac{c}{12}(n^3-n)\delta_{n+m}~,\nonumber\\
[L_n,P_m]&=-mP_{n+m}~,\nonumber\\
[P_n,P_m]&=-\frac{k}{2}n\delta_{n+m}~\,. \label{algebra}
 \end{align}
 This asymptotic symmetry algebra is identified with the symmetry algebra of the holographic dual warped CFT$_2$. The holographic map between the boundary field theory parameters ($c,k$) and bulk gravity action parameters ($G_N, l, \nu$) is \cite{Henneaux:2011hv},
	\begin{align}
c=\frac{5 \nu ^2+3}{\nu  \left(\nu ^2+3\right)} \frac{l}{G},\,\,\, k = \frac{\nu^2+3}{6\nu} \frac{l}{G}. \label{Holomap}
	\end{align}

Thus the final expression for complexity of warped CFT dual to spacelike warped AdS$_3$ is
	\begin{align} 
	    \mathcal{C}&= \tilde{c}\,\, \frac{L_x}{\epsilon},\label{swadsC}
	\end{align}
where the parameter $\tilde{c}$ is a rather elaborate function of the symmetry algebra parameters $c,k$,
\begin{equation} 
\tilde{c} = \frac{5^{\frac{5}{2}}}{2^{\frac{11}{2}}3^{\frac{1}{2}}}\,k\,\left(\frac{3}{5}-\sqrt{1-\frac{8c}{25k}}\right) \left(1+\sqrt{1-\frac{8c}{25k}}\right)^{\frac{3}{2}}. \label{ctildel}
\end{equation}
for $\nu<1.3416$ while for $\nu>1.3416$,
\begin{equation} 
\tilde{c} = \frac{5^{\frac{5}{2}}}{2^{\frac{11}{2}}3^{\frac{1}{2}}}\,k\,\left(\frac{3}{5}+\sqrt{1-\frac{8c}{25k}}\right) \left(1-\sqrt{1-\frac{8c}{25k}}\right)^{\frac{3}{2}}. \label{ctildem}
\end{equation}
The holographic complexity expression of the WCFT dual to spacelike WAdS$_3$  (\eqref{swadsC}, \eqref{ctildel} \eqref{ctildem}) has the following features of note,
	\begin{itemize}
	    \item Complexity scales extensively with the number of lattice sites, i.e. system size in units of the UV cutoff, $\mathcal{C}\propto L_x/\epsilon$ (here since the WCFT/system is spatially extended in one dimensions), much like that of a local field theory CFT$_2$. This is a bit counterintuitive since the WCFT is understood to be a nonlocal theory.
	    \item Unlike in the case of the WCFTs dual to timelike or null WAdS$_3$ case, there is no restriction of the UV cutoff $\epsilon$ on the warping parameter $k/c$. This affirms the fact that the dual WCFT to spacelike WAdS is a unitary UV complete theory.
	    \item In contrast to local CFT$_2$, for which the holographic complexity is proportional to the Virasoro central charge $c$, in the case of the WCFT$_2$ it is proportional to $\tilde{c}$, which is a complicated function of the Virasoro central charge and the $U(1)$ Kac-Moody level. We note that the for $\nu^2\in \mathbb{R},$ one restricts the range of the parameters $c,k$ to the domain
	    \begin{equation}
	    \frac{c}{k}\leq\frac{25}{8}.
	    \end{equation}
	    So there is no way to set $k\rightarrow0$ while keeping $c$ finite.
	    \item Finally one can check that setting $\nu^2 =1$ in \eqref{CV} or equivalently by setting $$c=\frac{2l}{G},\,k=\frac{2l}{3G}$$ in \eqref{ctildel} one recovers the pure AdS complexity\footnote{Incidentally, one could perhaps attempt to extract the volume complexity of spacelike WAdS$_3$ by taking the zero temperature (and zero angular velocity) limit of the holographic volume complexity expression for WAdS$_3$ black holes obtained in \cite{Auzzi:2019fnp}. However that volume complexity expression does not reduce to the pure AdS volume complexity when one sets $\nu^2=1$, in fact the divergences in volume complexity \emph{vanish entirely} in the unwarped pure AdS limit by setting $\nu^2=1$ and $M=0$ in Eq. 4.6 of \eqref{ctildel}.}.
	\end{itemize}

	\section{Circuit complexity for warped CFTs}\label{KacMoody}
In this section we compute the circuit complexity for dual warped conformal field theory using the approach as outlined in \cite{Caputa:2018kdj,Flory:2020eot, Erdmenger:2020sup}. This will allow us to compare and contrast the field theory based circuit complexity using available techniques to the holographic results of the last section. In general such a direct comparison of field theory and holographic results are rare, WAdS/WCFT complexity provides  us yet another opportunity.\\\\
\textit{Symmetry generators and their transformations:}\\\\
As discussed in \cite{Hofman:2011zj,Detournay:2012pc}, the Lorentzian theory has a global $SL(2, R)_R \times U(1)_L$ invariance. Furthermore, on the plane the algebra is defined by the commutators of the following operators \cite{Hofman:2011zj,Detournay:2012pc},
	\begin{align}
T_{\zeta}&=-\frac{1}{2\pi}\int dx^-\, \zeta (x^-)T(x^-), & P_{\chi}&= -\frac{1}{2\pi}\int dx^-\, \chi(x^-)P(x^-).
 \end{align}

The right moving and left moving modes are associated with $x^{-}$ and $x^{+}$
     respectively and $T(x^{-})$, $P(x^{-})$ are the local operators (the stress-tensor and momentum operator) on the plane. The ground state of the theory is invariant under the action of these symmetry generators.\par
     To get an insight about the algebra, let us take an concrete example. If we go from a Lorentzian plane $(x^+,x^-)$ to a  Lorentzian cylinder using the coordinate transformations $x^-=e^{i\phi}$ and choose the test functions 
     \begin{align}
     \begin{split}
    & \zeta(x^{-})=\zeta_n= (x^{-})^n= e^{i\,n\,\phi},\\&
     \chi(x^{-})=\chi_n= (x^{-})^n= e^{i\,n\,\phi},
     \end{split}
     \end{align}
then following \cite{Hofman:2011zj,Detournay:2012pc} one can show that that Fourier modes satisfy Virasoro-Kac-Moody algebra mentioned in (\ref{algebra}) with central charge $c$ and the Kac-Moody level $k$ after the following identification,
\begin{align}
    \begin{split}
   L_{n}=i\,T_{\zeta_{2n+1}},\quad P_{n}=P_{\chi_{n}}.     
    \end{split}
\end{align}
Note that, the $T(x^{-})$ generates infinitesimal coordinate transformation for the coordinate $x^{-}.$ On the other hand $P(x^{-})$ generates the infinitesimal gauge transformations in the gauge bundle parametrized by $x^{+}.$  Following \cite{Hofman:2011zj,Detournay:2012pc} we can write down the following transformation rules for $T(x^{-})$ and $P(x^{-})$
\begin{align}
    \begin{split} \label{coord1}
   & T'(w^{-})=f'^2\, T(x^{-})+\frac{c}{12}\{f,w^{-}\}+f'g'P(x^{-})-\frac{k}{4} g'^2,\\&
P'(w^{-})= f' P(x^{-})-\frac{k}{2}g',
\end{split}
\end{align}
where,  $f, g$ are two arbitrary functions and $f'=\frac{\partial f (w^{-})}{\partial w^{-}}, g'=\frac{\partial g(w^{-})}{\partial w^{-}}.$ Also we have used the fact that the finite transformations for the coordinates  going from $(x^{-}, x^{+})$ to $(w^{-}, w^{+})$  is of the form mentioned in (\ref{coordtrans}). Also, we can identify the Schwarzian term as, $$\{f,w^{-}\}=\frac{f'''}{f'}-\frac{3}{2}\Big(\frac{f''}{f'}\Big)^2.$$
Now again going back to the case of mapping the theory defined on a  plane to a cylinder, using the (\ref{coord1}) we get,
\begin{align}
\begin{split} \label{relat}
    T^{\alpha}(\phi)&=-(x^{-})^2\, T(x^-)+\frac{c}{24}+i\,2\,\alpha\, x^-P(x^-)+ k\,\alpha^2,\\ 
    P^{\alpha}(\phi)&=i\,x^-P(x^-)+ k\,\alpha
\end{split}
\end{align}
where we have used the following coordinate transformations, 
\begin{align}
x^{-}=e^{i\,\phi},\quad x^{+}=t+ 2\,\alpha\,\phi.\label{trans}
\end{align}
$\alpha$ is an arbitrary tilt \cite{Hofman:2011zj,Detournay:2012pc}.
The Fourier modes for $ T^{\alpha}(\phi)$ and $P^{\alpha}(\phi)$ on the cylinder is defined as, 
\begin{align}
    P_n^{\alpha}&=-\frac{1}{2\pi}\int d\phi\, P^{\alpha}(\phi)e^{in\phi} ~,& L_n^{\alpha}&=-\frac{1}{2\pi}\int d\phi\,T^{\alpha}(\phi)e^{in\phi}~.
\end{align}
Then using the (\ref{relat}) we can relate the Fourier modes on the cylinder with those on the planes in the following way, 
\begin{align}
    \begin{split}
        L^{\alpha}_n &= L_n+2\,\alpha\,P_n -\Big(k\,\alpha^2 +\frac{c}{24}\Big)\,\delta_n,\\
        P^{\alpha}_n&= P_n - k\,\alpha\,\delta_n,
    \end{split}
\end{align}
where $L_n$ and $P_n$ are the Fourier modes defined on the plane.\par We need to know one more thing before we proceed further. We will be requiring to know the expectation values of $T^{\alpha}(\phi)$ and $P^{\alpha}(\phi)$ with respect some primary states.\\\\
\textit{Complexity measure for symmetry groups:}\\\\
Now we use the method of \cite{Caputa:2018kdj} to compute the circuit complexity. In \cite{Caputa:2018kdj} authors have adapted the methods for computing circuit complexity \cite{2006Sci...311.1133N} for conformal field theory. We primarily follow their approach. Starting from a suitable reference state $|\psi_{R}\rangle$ we can go a target state $|\psi_{T}\rangle$ by acting the reference state by a unitary operator $$U(\tau)=\overleftarrow{\mathcal{P}}\text{exp}\left(-i\, \int_0^{\tau} H(\tau') d\tau'\right).$$ At $\tau=0$ this $U(\tau)$ becomes identity operator so that we get the reference matrix. Then,
\begin{align}
    |\psi_{T}\rangle = U(\tau=T) |\psi_{R}\rangle.
\end{align}
Here we have assumed that we will reach the target state after $\tau=T$ time. The Hermitian operator $H(\tau)$ is composed of a set of gates that satisfy a closed algebra and form a group. $\overleftarrow{\mathcal{P}}$ represents the path ordering as these gates do not commute in general. In \cite{Caputa:2018kdj}, the authors following the arguments of \cite{Magan:2018nmu} focuses on the symmetry group. Hence the gates are generated by the symmetry generators. This method has been used to compute circuit complexity for Virasoro and Kac-Moody groups \cite{Erdmenger:2020sup,Flory:2020eot}. \par

Using appropriate representation we can identify the instantaneous gates $Q(\tau')=-i\, H(\tau')$ as,
\begin{align} \label{gate1}    Q(\tau')=\frac{1}{2\pi}\int dx \,\epsilon(\tau', x) J(x),
\end{align}

where $J(x)$ is the conserved current and $\epsilon(\tau',x)$ is the control functions which counts how many times the particular generators have been acted at a given time $\tau'.$ One can view the circuit as a path on the underlying group manifold connecting two given points. For infinitesimally close points along the path we can write down the following, 
\begin{align} \label{tautrans}
    U(\tau+d\tau)= e^{-Q(\tau) d\tau} U(\tau). 
\end{align}

We also we need to relate this control function with the group elements to compute the circuit complexity. This can be done by noting the fact that under the symmetry transformations we can write the following for the group element $g(\tau,x),$
\begin{align} 
    g(\tau+d\tau,x)=e^{\epsilon(\tau, x) d\tau} g(\tau,x),\label{param}
\end{align}
Then we can expand this to the first order we can relate the control function with the derivative of the group element \cite{Caputa:2018kdj,Erdmenger:2020sup}. It can be easily seen that, $\epsilon(\tau,x)$ is nothing but the instantaneous velocity in the group space. \par
Finally we need to specify a suitable functional assigning computational cost to all of these symmetry transformations. In the original formulation by the Nielsen \cite{2006Sci...311.1133N}, typically one assign higher penalties for those gates which are `difficult' to construct. Here we will follow the approach of \cite{Caputa:2018kdj,Erdmenger:2020sup} to assign same cost all kind of symmetry transformations. Furthermore, following \cite{Caputa:2018kdj} we will define the cost functional by evaluating the gates $Q(\tau)$ in the instantaneous state at time $\tau.$ This is different from the Nielsen's original formulation.  For more details we  discussions on this interested readers are referred to \cite{Caputa:2018kdj,Erdmenger:2020sup}. We mainly use the following cost-functional, 
\begin{align} \label{costfunc}
    \mathcal{F}=|\mathrm{Tr}[\rho(\tau) Q(\tau)]|.
\end{align}
This is also known as ``one-norm" cost-functional. There are plethora of choices for cost-functional. For more details of possible choices for cost-functional interested readers are referred to \cite{2005quant.ph..2070N,Guo:2018kzl}.  The density matrix for the instantaneous state $\rho(\tau)$ in this can be generated from the initial density matrix $\rho_0$ by evolving with the unitary operator, $\rho(\tau)= U(\tau) \rho_0 U^{\dagger}(\tau).$ Then (\ref{costfunc}) can be re-written as,
\begin{align} \label{cost1}
    \mathcal{F}=|\mathrm{Tr}[\rho_0 \tilde Q(\tau)]|,
\end{align}
where, $\tilde Q(\tau)= U^{\dagger}(\tau) Q(\tau) U(\tau).$ Furthermore the total cost can be found by integrating over this cost-functional over the entire path connecting reference and target states.
\begin{align} \label{gate2}
    \mathcal{C}=\int d\tau \mathcal{F}=\frac{1}{2\pi}\, \int d\tau\, \Big|\int dx\, \epsilon(\tau,x)\langle \psi_R|U^{\dagger}(\tau) J(x) U(\tau)|\psi_R\rangle\Big |,
\end{align}
where we have used (\ref{gate1}) and (\ref{cost1}). Then we have to choose a suitable reference state and minimize (\ref{gate2}). Note that, (\ref{gate2}) is a functional of the group path $g(\tau).$ By minimizing it we are finding the shortest path. Also as each path corresponds a circuit, shortest path corresponds  to the optimal circuit. Finally, evaluating (\ref{gate2}) on this path will give us the complexity associated with the optimal circuit which will take us from a given reference state to a desired target state.\\\\
\textit{Virasoro-Kac-Moody Circuit:}\\\\
Armed with this discussion, now we turn our attention to the Virasoro-Kac-Moody symmetry group and compute the circuit complexity using the methods discussed above. We construct the unitary circuit solely using the gates generating Virasoro-Kac-Moody symmetry defined by,
\begin{align}
\begin{split}
    Q_T(t)&=\int_0^{2\pi}\frac{d\phi}{2\pi}\,\epsilon_1(t,\phi) T^{\alpha}(\phi) ,\\
    Q_P(t)&=\int_0^{2\pi}\frac{d\phi}{2\pi}\,\epsilon_2(t,\phi) P^{\alpha}(\phi) \\
    \end{split}\label{Fourier}
\end{align}
where, $T^{\alpha}(\phi)$ and $P^{\alpha}(\phi)$ are the stress-tensor and momentum operator defined on the cylinder (\ref{relat}). The quantum circuit then takes the following form,
\begin{align}
    U(\tau)= \overleftarrow{\mathcal{P}}\text{exp}\left[\int_0^{\tau}\Big(Q_T(\tau')+Q_P(\tau')\Big) d\tau'\right].
\end{align}
Next to compute the complexity functional (\ref{gate2}) we have to do the following:
\begin{itemize}
    \item We choose the reference state $|\psi_R\rangle$ as the primary state $|p, h\rangle $  of the underlying warped CFT \cite{Hofman:2011zj,Detournay:2012pc}. 
    \item To compute the following,
\begin{align} \label{trans1}
\tilde Q(\tau)= U^{\dagger}(\tau)\Big(Q_T(\tau)+Q_P(\tau)\Big) U(\tau).
\end{align}
we first note that, $U(\tau)$ is basically unitary representation of the symmetry group elements. Hence acting $U(\tau)$ on $Q_T(\tau)$ and $Q_P(\tau),$ amounts to transform $T^{\alpha}(\phi)$ and $P^{\alpha}(\phi)$ using the transformation rules mentioned in (\ref{coord1}). We  then get the following, 
\begin{align}
    \begin{split}
   &  U^{\dagger}(\tau) T^{\alpha} U(\tau)= f'(\tau,\phi)^2\,T^{\alpha}(\phi)+\frac{c}{12}\{f(\tau,\phi),\phi\}+f'(\tau,\phi)g'(\tau,\phi) P^{\alpha}(\phi)-\frac{k}{4} g'(\tau,\phi)^2,\\&
    U^{\dagger}(\tau) P^{\alpha} U(\tau)=f'(\tau,\phi) P^{\alpha}(\phi)-\frac{k}{2}g'(\tau,\phi). \label{UTU-UTP}
    \end{split}
\end{align}
Here, for a given path $\tau$ in the group manifold, $f$ is the diffeomorphism on the circle just like the Virasoro case \cite{Caputa:2018kdj,Erdmenger:2020sup} and $g$ provides a translation along $\tau$ for given $\phi.$ Note that in contrast to \cite{Caputa:2018kdj,Erdmenger:2020sup} we are using the notation $f,g$ instead of $F,G$ and to represent the inverse diffeomorphism and vice versa. Our notation is more in line with the original literature in the context of coadjoint orbit action in 2D gravity \cite{Alekseev:1988ce}.
\item Also we relate the the control functions $\epsilon_{1,2}(\tau,\phi)$ with the group parameters. Note that, we can identify it as the instantaneous velocity in the group space from (\ref{param}) just like the case of Virasoro \cite{Caputa:2018kdj,Erdmenger:2020sup}. 
\begin{align}
   \epsilon_1(\tau,\phi)=-\frac{\dot{f}(\tau,\phi)}{f'(\tau,\phi)}\,,\label{e1}
\end{align}
\begin{equation}
\epsilon_2(\tau,\phi)=\frac{\dot{g}(\tau,\phi)\,f'(\tau,\phi)-g'(\tau,\phi)\,\dot{f}(\tau,\phi)}{f'(\tau,\phi)}\,.\label{e2}
\end{equation}
We have denoted the $\tau$ and $\phi$ derivative as by ``$(\,\dot{}\,)$" and $(\, '\, )$ and respectively. Details of the derivation can be found in appendix \ref{App.C}. At this point we note that, $\epsilon_1(\tau,\phi)$ depends only on $f(\tau,\phi)$ and its derivative just like the Virasoro case \cite{Caputa:2017yrh, Erdmenger:2020sup}.  
 \item Finally we replace the expectation values of $T^{\alpha}(\phi)$ and $P^{\alpha}(\phi)$ with respect to the primary states in the complexity functional (\ref{gate2}) by the relations mentioned in (\ref{expec}).
\end{itemize}
\textit{Complexity Functional:}
Using \eqref{UTU-UTP}, \eqref{e1} \eqref{e2}, we arrive at the form of the complexity functional,
\begin{align}
\begin{split} \label{action}
   & \mathcal{C}=\frac{1}{2\pi}\int_{0}^{T} d\tau\,\int_0^{2\pi} d\phi\, \Big|\epsilon_1\, \langle \psi_R| U^{\dagger}(\tau) T^{\alpha} U(\tau)|\psi_R\rangle+\epsilon_2\, \langle \psi_R|U^{\dagger}(\tau) P^{\alpha} U(\tau)|\psi_R\rangle \Big|\,\\&
   = \frac{1}{2\pi}\int_{0}^{T} d\tau\,\int_0^{2\pi} d\phi\, \Big|-\dot{f}f'\,T_0-\frac{c}{12}\frac{\dot{f}}{f'}\{f,\phi\}-2\,\dot{f}\, g'\, P_0+\frac{3\,k}{4}\frac{\dot{f}}{f'} g'^2+\dot{g}\, f'\,\,P_0-\frac{k}{2}\,\dot{g}\, \,g'\Big|\,.
    \end{split}
\end{align}
Here we have defined $T_0=\langle h| T^{\alpha}(f,g) |h \rangle$ and $P_0=\langle h| T^{\alpha}(f,g)|h \rangle$, and we have chosen the the primary state $|h\rangle$ as the reference state $|\psi_R\rangle$. Now, before we extremize this complexity functional we switch to the inverse functions $F,G$ (refer to Sec. \ref{App.C}) by switching variables $\phi\rightarrow f(\tau, \phi)$.
Thus, we arrive at the complexity functional form,
\begin{align}
\begin{split}
\mathcal{C}=\frac{1}{2\pi}\int_0^{T} d\tau \int_0^{2\pi} d\phi\left|\frac{\dot{F}}{F'}T_0-\frac{c}{12}\frac{\dot{F}}{F'}\{F,\phi\}-\frac{k}{4} \frac{\dot{F}\, G'^2}{F'} + \frac{G'\dot{F}}{F'} P_0+\dot{G} P_0-\frac{k}{2} \dot{G}\,G'\right|. \label{complexity functional}
    \end{split}
\end{align}
Extremizing this complexity functional leads to the Euler-Lagrange equations for $F(\tau,\phi)$:
\begin{align}
\partial_{\tau}\left[\frac{T_{0}}{F'}-\frac{c}{12}\frac{\left\{ F,\phi\right\} }{F'\,^{2}}-\frac{k}{4}\frac{G'\,^{2}}{F'}+\frac{G'}{F'}P_{0}\right]+\partial_{\phi}\left[\frac{\dot{F}}{F'\,^{2}}\left\{ \frac{c}{6}\left(\frac{F'''}{F'}-\frac{9}{2}\frac{F''\,^{2}}{F'\,^{2}}\right)-T_{0}\right\} \right]\nonumber \\
+\frac{c}{4}\partial_{\phi}^{2}\left(\frac{\dot{F}F''F'''}{F'^{3}}\right)+\frac{c}{12}\partial_{\phi}^{3}\left(\frac{\dot{F}}{F'^{2}}\right) & =0,\label{eq: EOM for F}
\end{align}
and for $G(\tau,\phi)$:
\begin{align}
    k\, \dot{G}'+\frac{k}{2}\frac{\dot{F}}{F'}G''+\left(\frac{\dot{F}}{F'}\right)'\Big(\frac{k}{2}G'-P_0\Big)=0. \label{EOM for G}
\end{align}
The arguments, $(\tau,\phi)$ of all the functions have been suppressed to reduce clutter of notation.\\

These are a pair of coupled nonlinear partial differential equations, of high (cubic) order in derivatives of $\phi$, and it is not a priori obvious what are the consistent boundary and initial data on $F,\dot{F}, F', F'',\dot{G},G'$ which will lead to the existence of an unique solution. The questions of consistent initial and boundary data, existence, uniqueness, boundedness of the solution etc of this equation can perhaps be taken up in a separate work. For now we content ourselves by arriving at a solution by plain guessing, since for the purpose of this paper, \emph{any} solution will allow us to make an estimate of the circuit complexity and afford a comparison with the holographic complexity computed in Sec. \ref{CVV}. By simple inspection, the most obvious solution are when $\dot{F},F'$ are constants, say $\dot{F}=k_1,\,F'=k_2$. Then the equation for $G$, $\eqref{EOM for G}$  yields,
\begin{equation}
G(\tau,\phi)= c_1(\tau)+c_2\left(\tau-2 k_2\phi/k_1\right)
\end{equation}
where $c_1, c_2$ are two arbitrary functions. Plugging this in the equation for $F$, $\eqref{eq: EOM for F}$ then implies either $G'=\frac{2}{k} P_0$ or $\dot{G}'=0$. The first choice determines $c_2$ to be,
\begin{equation*}
    c_2\Big(\tau-\frac{2k_2}{k_1}\phi\Big)=-\frac{k_1\,P_0}{k\,k_2}\Big(\tau-\frac{2k_2}{k_1}\phi\Big)+c_3
\end{equation*}
and thus,
\begin{equation}
    G(\tau,\phi)=c_1(\tau)+\frac{P_0}{k}\frac{k_1}{k_2}\Big(\frac{2 k_2}{k_1}\phi-\tau\Big)+c_3
\end{equation}
The second choice leads to the condition $c_2''(\tau-2k_2\phi/k_1)=0$, or,
\begin{equation*}
c_2(\tau,\phi) = k_3(2k_2\phi/k_1-\tau)+c_3.
\end{equation*}
where $k_3,c_3$ are constants. Thus the other solution is,
\begin{equation}
G(\tau,\phi) = c_1(\tau)+k_3(2k_2\phi/k_1-\tau)+c_3.
\end{equation}
One must have $G(\tau,\phi)$ only as a function of $\phi$ at $\tau=0$, which implies for either of the two solutions for $G$,
\begin{equation}
    c_1(0)=-c_3
\end{equation}
This will then give,
\begin{equation}
    G(\tau,\phi)=c_1(\tau)-c_1(0)+\gamma \Big(\phi-\frac{k_1}{2\,k_2}\tau\Big)
\end{equation}
where $\gamma = \frac{2\,P_0}{k}$ for the first solution and $\gamma=\frac{2\,k_3\,k_2}{k_1}$ for the second solution. Now if we impose an initial condition on $G$ so that it reduces to the phane-to-cylinder transformation at $\tau=0$, namely, $\eqref{trans}$, then we must have $\gamma = 2\alpha$.  However, $k_1, c_1(\tau)$ still remain arbitrary. From the periodicity condition on $F(\tau,\phi)$, namely $F(\tau,\phi+2\pi) = F(\tau,\phi)+2\pi$, we have $k_2=F'=1$. The solution to $F$, then assumes the form $F(\tau,\phi)=k_1 \tau+\phi+\text{const}$. The constant has to be set to zero, to satisfy the intitial condition, $F(0,\phi)=\phi$. Thus we have the final form of the solution,
\begin{align}
   & G(\tau,\phi)=c_1(\tau)-c_1(0)+2\alpha \Big(\phi-\frac{k_1}{2}\tau\Big),\\&
    F(\tau,\phi)=k_1\,\tau+\phi. \label{FG sol}
\end{align}
We will specialize to the vacuum state on the cylinder, for which $h=0$, whereby \cite{Hofman:2011zj,Detournay:2012pc} 
\begin{align}
    \begin{split} 
        T^{\alpha}(\phi) =  k \alpha^2+\frac{c}{24} ,\quad P^{\alpha}(\phi)= k \alpha. 
    \end{split} \label{expec}
\end{align}
Substituting the solutions $\eqref{FG sol}$ in $\eqref{complexity functional}$, and using the expectation values $\eqref{expec}$, we obtain the final expression for the complexity,
\begin{align}
\begin{split} \label{finalanswer}
     \mathcal{C}& = T\, k_1\, k\,\Big(2\alpha^2+\frac{c}{24\,k}\Big).
\end{split}
\end{align}
Now, before we extremize this complexity functional we switch to the inverse functions $F,G$ by switching variables $\phi\rightarrow f(\tau, \phi)$. Before we end this section, there are a few comments in order. 
 \begin{itemize}
     \item Among various other choices for complexity functional, there is one other commonly use functional which can be defined as, $\mathcal{F}=\sqrt{-\mathrm{Tr}[\rho_0\, \tilde Q^2(\tau)]}.$ For our case then the complexity will take the following form, 
  \begin{align}
      \mathcal{C}=\int_0^{T} d\tau \sqrt{-\mathrm{Tr}\Big[\rho_0\,(\tilde Q^2_T
      (\tau)+\tilde Q^2_P(\tau))\Big]}.
  \end{align}
  It has been argued in \cite{Caputa:2018kdj} this will give the same complexity as before in large $c$ limit. This is indeed the case for Virasoro symmetry group. Unless we take large $c$ limit we can not expect that this complexity functional will give the same value as what we have quoted in \eqref{action}.
 
  \item 
  For our case it is natural that we have to take a large $k$ limit as well  since the leading semiclassical limit is defined by $l/G_N \gg1$ which mandates large $c$, \emph{as well as} large $k$ since $\nu$ is order one). The final expression of the symmetry gate complexity \eqref{finalanswer} evidently displays a leading contribution of order $k$, with a subleading contribution of order $c/k$. Thus it is similar to the result obtained from the holography, which is also proportional to the $k$ in this limit as evident from (\ref{swadsC}), (\ref{ctildel}) and (\ref{ctildem}).
  \item Last but the not the least, unlike the gravity result, we do not get any UV cut-off dependence (short distance singularities) in the complexity! In fact this is the case for Virasoro group as well, as noted in the earlier literature \cite{Erdmenger:2020sup,Flory:2020eot}.
  \end{itemize}

\section{Discussions} \label{discussion}

WAdS$_3$/WCFT$_2$ duality allows us to explore holography beyond asymptotically AdS spacetimes. WCFTs are nonlocal quantum field theories characterized by the infinite dimensional symmetry algebra, namely the Virasoro-Kac-Moody current algebra. In this work we have probed WCFTs by means of circuit complexity, a novel tool which has traditionally been used in quantum information and computation theory, but has gained importance in black hole physics and holography of late. In particular we studied the WCFT complexity in two independent schemes. First is the holographic volume complexity scheme and the other is the recently proposed circuit complexity based on circuits constructed purely by means of unitary gates which are the Kac-Moody symmetry transformations. We mainly focused on WCFTs which are putative duals of spacelike warped AdS$_3$ since the WCFTs dual to timelike or null warped AdS$_3$ are not expected to have unitary UV completion. (While discussing holographic complexity, we did discuss the timelike warped AdS$_3$ case just to illustrate the point that the complexity expression becomes nonanalytic and develops cuts when the UV cutoff is made arbitrarily small signaling UV incompleteness of the dual WCFT). For spacelike warped AdS$_3$ case, the dual WCFT$_2$ holographic complexity turns out to be linearly divergent. This is rather counterintuitive because such linear divergences are expected for \emph{local} CFT$_2$ while WCFTs are nonlocal field theories. However, such a trend has been true for other observables like WCFT entanglement entropy \cite{Castro:2015csg, Anninos:2013nja} which does display logarithmic divergence characteristic of local CFT$_2$s. The coefficient of the complexity linear divergence for CFT$_2$ is the central charge (up to a numerical factor), while for the case of WCFT$_2$ we see that this coefficient is a rather elaborate function of the Virasoro central charge, $c$ as well as the $U(1)$ Kac-Moody level number, $k$ and the complexity only makes sense for the range of parameters in the domain $c/k \leq 25/8$. So there is no simple way to take a large $c$ limit while keeping $k$ fixed, in fact one has take both $c, k$ large while maintaining $c/k \leq 25/8$. However, one can take $k$ large while keeping the ration $c/k$ fixed, and in this limit, the holographic complexity has a leading behavior proportional to $k$. From the holographic standpoint, one might think of employing other schemes such as the action complexity or some other alternative proposals \cite{Belin:2021bga}. However we recall that for the case of CFT$_2$ dual to pure AdS$_3$, the action complexity vanishes due to dimensional accident (in arbitrary boundary spacetime dimensions, say $d$, the action complexity is proportional to a factor $\ln(d-1)$) \cite{Reynolds:2016rvl}. Analogous vanishing of the action complexity has also been observed in the null warped AdS$_3$ \cite{Katoch:2022hdf} while being studied in the broader context of holographic complexity of little string theories \cite{Chakraborty:2020fpt}.  So we do not pursue this direction here in this work. See \cite{Ghodrati:2017roz} for a calculation of the divergence-free time-rate of action-complexity growth in warped AdS black hole geometries.

Next we looked at the circuit complexity of WCFT based on a proposal \cite{Erdmenger:2020sup} which advocates the use of the unitary gates corresponding to (exponentiating) Kac-Moody symmetry generators. Note that, the complexity functional mentioned in (\ref{action}) is not actually geometric action functional supported on the coadjoint orbits of the Virasoro-Kac-Moody symmetry group \footnote{This is also true for the Virasoro case if we simply follow the approach of \cite{Caputa:2018kdj}.}. In \cite{Erdmenger:2020sup}, a modification of the proposal advocated in \cite{Caputa:2018kdj} has been given. It will be interesting to use that modification to obtain a complexity functional for our case which will be same as warped coadjoint orbit action. However, extremizing the complexity functional (\ref{action}) leads to a pair of highly nonlinear coupled PDEs which appear to be quite intractable. But by simple inspection, we find can a special solution where the derivatives $F',\dot{F}$ are constants. This choice yields a very simple looking expression for the circuit complexity, one which depends on the symmetry parameters $c$ and $k$, and the path length $T$ between the reference and the target state. Of course it would be ideal if one could somehow arrive at a more general solution of the complexity functional extremization conditions, and figure out the most generic dependence on the parameters $c, k$. At this point it is unclear what are the consistent boundary conditions which will lead to a well-behaved solution of (\ref{eq: EOM for F}). Perhaps one may try to consider to particular limits involving $c$ and $k$ to simplify (\ref{eq: EOM for F}) and get a well behaved solution. We leave that general analysis for future work. Since the complexity functional is proportional to the length of the path in the group manifold and not physical space, there is no dependence on the system size. Also, by construction this complexity functional is tolerance free and free from UV divergences. Thus perhaps one cannot perform a direct comparison of the resulting expressions of the two schemes of holographic complexity and circuit complexity employed here, apart from the fact that \emph{both} are proportional to $k$ in the leading order. This absence of UV divergences might be related to the choice of the reference state. The reference state considered in this paper is a primary state. However, one might think of trying to setup this computation in the spirit of \cite{Jefferson} where one start from a unentangled reference state. This will be an interesting avenue to pursue in future. Nevertheless, at this point, these are only speculations. Further investigations are needed to understand the absence of UV-divergence from the complexity computed using this group-theoretic approach.

In Nielsen's original formalism \cite{2006Sci...311.1133N,2005quant.ph..2070N, Bhattacharyya:2019kvj},  this  $\mathcal{C}$ in (\ref{gate2}), can furthermore be related to the number of gates, which in turn then make the complexity dependent on the system size. One first needs to perform that analysis for this case in order to relate the  $\mathcal{C}$  that we have computed in (\ref{action}) with the number of gates constituting the circuit. Also, one important thing that we have to keep in mind is that the penalty factor played an important role in such analysis. In our case, we have not penalized any gates. We leave these important issues for future investigations. Apart from this, it will be also interesting  to investigate circuit complexity using other methods eg Fubini-Study, path integral approach along the lines of \cite{Erdmenger:2021wzc,Flory:2020eot,Caputa:2017yrh,Bhattacharyya:2018wym,Camargo:2019isp}.  Last but the not the least, it will be worthwhile to investigate operator complexity related with the Hamiltonian evolution. In that context an useful approach might be to consider recently proposed `Krylov complexity' \cite{Dymarsky:2021bjq,Banerjee:2022ime} for our case. Again we hope to report on this issue in near future.

\section*{Acknowledgements}

A.B is supported by Mathematical Research Impact Centric Support Grant (MTR/2021/000490), Start-Up Research Grant (SRG/2020/001380) by the Department of Science and Technology Science and Engineering Research Board (India) and Relevant Research Project grant (202011BRE03RP06633-BRNS) by the Board Of Research In Nuclear Sciences (BRNS), Department of Atomic Energy, India. The work of GK was supported partly by a Senior Research Fellowship (SRF) from the Ministry of Education (MoE), Govt. of India and partly from the RDF fund of SR: RDF/IITH/F171/SR. The work of SR is supported by the IIT Hyderabad seed grant SG/IITH/F171/2016-17/SG-47.

\appendix
\section{Review of timelike WAdS$_3$}\label{App.A}
Our starting point is the black string metric, equation (4.10) of \cite{Song:2016gtd}
	\begin{align}
	ds'^2&=l^2 \left(\frac{(1+\lambda^2A^2)}{4 \left(r^2-A^2 B^2\right)}dr^2+A^2 dv^2+(B^2 (1+\lambda^2A^2)-\lambda^2r^2)du^2+2r du dv\right)~,\label{blackstring}
	\end{align}
	where, the non-compact event horizon is located at $r_h=AB$. We will rewrite the above metric in order to arrive at timelike WAdS$_3$ metric. To this end let's separately write the $\lambda$ independent unwarped part of the metric as
    \begin{align*}
		ds_0^2&= \frac{\text{$dr$}^2}{4 \left(r^2-A^2 B^2\right)}+A^2 dv^2+B^2 du^2+2r\, du dv\,, \nonumber
	\end{align*}
	where we have taken $l=1$. If we further perform the following set of coordinate changes and parameter redefinitions,
\[
du=dx+dt,\quad dv=dx-dt,\quad r=\frac{1}{2}\left(\rho^{2}-\frac{a^{2}+b^{2}}{2}\right),
\]
\[
A=\frac{a+b}{2},\quad B=\frac{a-b}{2}.
\]
	then the black string metric \eqref{blackstring} turns into	
	\begin{align*}
		ds_0^2&=\frac{\text{d$\rho $}^2  \rho ^2}{\left(\rho ^2-a^2\right) \left(\rho ^2-b^2\right)}-\frac{ \left(\rho ^2-a^2\right) \left(\rho ^2-b^2\right)}{\rho ^2}dt^2+\rho ^2 \left(dx -\frac{ a b}{\rho ^2}\,dt\right)^2~.\nonumber
	\end{align*}	
	This is evidently the metric of BTZ spacetime in disguise with horizons at $a \,\text{and}\, b.$	In this setup, $A$ is related to the level of Kac-Moody algebra.
	As we can see that in order to obtain a black hole free background by making horizon disappear simply amounts to taking $a=0=b$ (i.e. vanishing Kac-Moody level).
This choice simplifies the metric to the form
	\begin{align*}
	ds_0^2&= \frac{\text{$dr$}^2}{4 r^2}+2\,r\, dudv ~,
	\end{align*} 
	thus we recover the pure $AdS_3$  metric
	\begin{align*}
ds_0^2&=\frac{\text{d$\rho $}^2}{\rho ^2}+(-\text{d$t$}^2 +\text{$dx$}^2) \rho ^2,
	\end{align*}	
	which we immediately recognise to be the metric of the Poincare patch of the $AdS_3$.	
After taking $\rho\to \frac{1}{z}$, further simplifies the metric to
	\begin{align*}
ds_0^2=\frac{dz^2-dt^2+dx^2}{z^2}~.
	\end{align*}
Let's turn our attention towards the warped part of the metric . After plugging in $A=0=B$ in (\ref{blackstring}) and carrying out the exact same transformations as above, leads us to the warped portion of the metric 
	\begin{align*}
ds_{\lambda}^2=	-\lambda ^2\frac{ (dt+dx)^2}{4 z^4}~.
	\end{align*}
Hence, the required timelike  $WAdS_3$ metric we work with in section (\ref{CV}) takes the following form
	\begin{align}
ds^2=\frac{dz^2-dt^2+dx^2}{z^2}	-\lambda ^2\frac{ (dt+dx)^2}{4 z^4}~.\label{TWAdS}
	\end{align}
\section{Maximum volume spacelike hypersurfaces}\label{App.B}
 The (regulated) volume for the $t=T_0$ slice is,
\begin{equation}
\mathcal{V}=a\:\int_{0}^{\frac{\pi}{2}-\delta}\frac{d\theta}{\cos\theta}=a\:\left[-\ln\left(\frac{\delta}{2}\right)-\frac{\delta^{2}}{12}-\frac{7\delta^{4}}{1440}+O\left(\delta^{6}\right)\right].\label{eq:maxima}
\end{equation}
Despite the minus sign the leading log divergent contribution is positive
definite because $\delta\ll\pi/2$, so $\delta/2\ll\pi/4\ll1$. Note that the UV regulator $\delta$ is related to the UV regulator $\epsilon$ by, 
\begin{equation}
\frac{1}{\epsilon} = -\ln\left(\frac{\delta}{2}\right)
\end{equation}
courtesy the diffeomorphism $\tan{\theta}=\sinh{\rho}$. One can compare the volume of this spacelike slice to the volume for the two other (family) of solutions for which
\[
\dot{t}(\theta)=\pm\frac{\cos\theta}{\sqrt{e^{2c_{1}}+\cos^{2}\theta}},
\]
where now we have redefined $c_1\rightarrow \exp{c_1}$, so that now $c_1 \in \mathbb{R}$. If one works out the normals to the above hypersurface(s), one finds
\begin{align}
   n_{t}&=1 &n_{\theta}&=\mp\frac{\cos \theta }{\sqrt{c_1+\cos ^2(\theta )}}& n_u&=0~, \label{unnornmalized}
\end{align}
with (regulated) norm at the conformal boundary\footnote{Clearly as $\theta\rightarrow\pi/2$ the metric diverges due to the overall factor of $\cos^{-2}\theta$, and the norm of the unnormalized normal vector \eqref{unnornmalized} vanishes. So one has to work with the normalized normal vector. Alternatively, one can use the metric without the $\cos^{-2}\theta$ factor and compute the norm.}
\begin{align*}
    n^2\propto\left(4 e^{2 c_1}+3\right) \nu ^2+8 \sqrt{e^{2c_1}} \nu ^2+3~.
\end{align*}
Evidently, for $e^{c_1}>1+\frac{\sqrt{\nu ^2+3}}{2 \nu}$ the normal is timelike ($n^2<0$) and the solutions \eqref{null hypersurfaces} represent spacelike hypersurfaces. The regulated volume for these one-parameter family of spacelike hypersurfaces are given by the same
expression,
\begin{align}
\mathcal{V} & =a\:\int_{0}^{\frac{\pi}{2}-\delta}\frac{d\theta}{\cos\theta}\sqrt{\frac{1}{1+e^{-2c_{1}}\cos^{2}\theta}}\nonumber \\
 & =a\:\tanh^{-1}\left(\frac{\cos\delta}{\sqrt{1+e^{-2c_{1}}\sin^{2}\delta}}\right)\nonumber \\
 & =-\frac{a}{2}\ln\left(\frac{\sqrt{1+e^{-2c_{1}}\sin^{2}\delta}-\cos\delta}{\sqrt{1+e^{-2c_{1}}\sin^{2}\delta}+\cos\delta}\right)\nonumber \\
 & =-a\ln\left(\frac{\delta}{2}\right){-\frac{a}{2}\ln\left(1+e^{-2c_{1}}\right)}-\left(1-3e^{-2c_{1}}\right)\frac{\delta^{2}}{12}+O\left(\delta^{4}\right).\label{eq: minima}
\end{align}
Evidently this solution has a lower volume is than the solution $t(\theta)=\text{constant}$
case (\ref{eq:maxima}) since the finite piece (indicated in red)
is negative definite for finite $c_1$. So the solution solution $t(\theta)=\text{constant}$
which we have worked with is a global maxima.

\section{Relating control functions with the group velocities}\label{App.C}
If we compose two warped conformal transformations,
\begin{equation*}
x^- \rightarrow x'^-= F_1(x^-),\,\, x^+\rightarrow x'^+= x^++G_1(x^-)
\end{equation*}
\begin{equation*}
x'^- \rightarrow {x''}^{-}= F_2(x'^-),\,\, x'^+ \rightarrow{x''}^+ =x'^+ + G_2(x'^-),
\end{equation*}
we get,
\begin{equation*}
{x}^-\rightarrow {x''}^-= F_2\left(F_1 (x^-)\right),\,\, x^+\rightarrow {x''}^+=x^++G_1 (x^-)+G_2 \left(F_1(x^-)\right).
\end{equation*}
So the group composition law is,
\begin{equation}
(F_2,G_2). (F_1,G_1) = (F_2\circ F_1,\,\,G_2\circ F_1+G_1)
\end{equation}
Using this transformation law in the equation for the group space circuit, namely
\begin{equation*}
\mathcal{G}(\tau+d\tau,\phi)= e^{\epsilon(\tau, \phi) d\tau}.\,\mathcal{G} (\tau,\phi)
\end{equation*}
where $\mathcal{G}=(F,G)$ is a group element and the infinitesimal transformation is defined by
\begin{equation*}
e^{\epsilon (\tau,\phi)} d\tau \equiv \mathds{1}+(\epsilon_1 (\tau,\phi) d\tau, \epsilon_2(\tau,\phi) d\tau).
\end{equation*}
We obtain,
\begin{equation}
\epsilon_1(\tau,F(\tau,\phi)) = \dot{F}(\tau, \phi),\,\, \epsilon_2(\tau,F(\tau,\phi)) =\dot{G}(\tau, \phi)\,.
\end{equation}
The solution for $\epsilon_1$ can be written in terms of the inverse $f(\tau, F(\tau, \sigma))=\sigma$,
\begin{equation}
\epsilon_1= -\frac{\dot{f}}{f'}.
\end{equation}
To solve for $\epsilon_2$ we need to introduce a function $g$, defined by $g(\tau, F(\tau,\sigma))= G(\tau,\sigma)$. This is the condition of the inverse of the transformation $(F,G)$. This then leads to form of the solution,
\begin{equation}
\epsilon_2=\dot{g}+g'\,\epsilon_1=\frac{\dot{g}\,f'-g'\,\dot{f}}{f'}.
\end{equation}\\
Here and henceforth, dot ($\,\,\dot{}$\,\,) and prime ($\,\,'\,\,$) denote partial derivatives wrt to $\tau$ and $\sigma$ respectively.


\bibliographystyle{utphysmodb}
\bibliography{ref}

\end{document}